\documentclass[%
reprint,
showpacs,
amsmath,amssymb,
aps,
pra,
]{revtex4-1}

\usepackage{graphicx}
\usepackage{dcolumn}
\usepackage{bm}
\usepackage{graphicx}
\usepackage{dcolumn}
\usepackage{bm}
\usepackage{bigints}
\usepackage{color}
\usepackage{amsmath,amssymb,graphicx}
\usepackage{float}
\usepackage{braket}

\begin{document}

\preprint{APS/123-QED}

\title{Design of a wavelength-tunable optical tweezer using a graded-index multimode optical fiber}

\author{Esmaeil Mobini}
\author{Arash Mafi}%
 \email{mafi@unm.edu}
\affiliation{Department of Physics \& Astronomy, University of New Mexico, Albuquerque, NM 87131, USA
             \\Center for High Technology Materials, University of New Mexico, Albuquerque, NM 87106, USA}%

\date{\today}

\begin{abstract}
A wavelength-tunable Optical Fiber Tweezer (OFT) based on a Graded Index Multimode Fiber (GIMF) with a flat endface is proposed.
It is shown that the design can support a trapping position which is far from the tip of the GIMF compared with other common
optical tweezing methods, hence reducing the possibility of a contact between the trapped particle and the fiber tip. Moreover,
because of the wavelength dependence of the GIMF design parameters such the Numerical Aperture (NA), the trapping position 
can become wavelength-dependent. Therefore, the trapping position can be tuned over a long range using a common wave-length tunable
laser. The proposed OFT differs from previous fiber-based demonstrations by using a flat-endface fiber making the fabrication and 
experiment quite easier than previously proposed tapered-endface OFTs.              
\end{abstract}
\maketitle
\section{Introduction}
Optical Tweezer has been extensively investigated since it was first developed by Ashkin~\textit {et al}~\cite{Ashkin:86,Keir,RodriguesRibeiro:15,Constable:93}.
They argued that it is possible to generate a strong trapping force in a region of high optical field intensity 
gradient and demonstrated an optical tweezer. In order to achieve the large intensity gradient, they used a high NA objective lens. 
A tweezer based on an objective lens is bulky and the tweezer does not have much flexibility to be moved over the samples. 
In order to address this problem, optical fiber tweezers have been under intense
study in recent years~\cite{Lyons,K.Tagucht}. 

In order to enhance the effective NA and hence the trapping efficiency of the optical fiber tweezer, 
Lyons~\textit {et al.}~\cite{Lyons} and Taguchi~\textit {et al.}~\cite{K.Tagucht} designed a tapered 
and lensed fiber tip geometry. However, the method resulted in a small trapping distance and required a very complex fabrication process.
In order to further enhance the effective NA and increase the trapping distance compared with the hemispherical or spherical lens designs at the
tip of the fiber, parabolic fiber tweezers were proposed in Refs.~\cite{Liu:06,Hongbao:Xin}. Theses optical fiber tweezers were fabricated 
using a single mode fiber (SMF) and resulted in an acceptable three-dimensional (3D) optical trap. 

More recently, Yuan~\textit {et al.}~\cite{Gong:13} proposed a fiber-based 
optical tweezer and argued that the self-focusing properties of the GIMF can result in a good 3D optical trapping as well as a large
trapping distance. The trapping distance is important for the design of a viable optical fiber tweezer because it is desirable to keep the
biological cells as far as possible from the tip of the fiber to reduce the possibility of a contact damage to the cells. 
A practical design of a GIMF-based tweezer was shown in Ref.~\cite{Gong:14} by Yuan~\textit {et al.} where the optical 
field was launched into the GIMF from an SMF and the launch condition could be customized by controlling a small gap between
the SMF and GIMF. The output tip of the GIMF in this paper was tapered to increase the trapping power. More recently, 
Zhang~\textit {et al.} Ref.~\cite{Zhang} demonstrated a flat endface GIMF optical fiber tweezer where the trapping distance could be 
changed by stretching the GIMF and therefore manipulating the relative phase relationship between the guided modes of the fiber.

In this manuscript, we present a detailed analytical description of the GIMF-based optical tweezers. Our results can explain the experimental
observations of previous research in this area, especially that of Zhang~\textit {et al.} Ref.~\cite{Zhang}. The analysis presented here
highlights the best practices to obtain a large trapping force as well as a large and controllable trapping distance. The formulation of the problem
is largely based on the previous work of our research group on multimodal interference in GIMFs and its device applications as a low-loss fiber 
adapter and a spectral filter~\cite{Mafi:11,Hofmann:12}. The novelty of this work, in addition to the rigorous analytical formulation,
is showing that the trapping distance can be controlled over a wide range by using a common laboratory wavelength-tunable laser.
 \par

\section{Radiation Forces}
In order to analyze the optical fiber tweezer and establish the notation, we briefly describe the nature of the optical forces involved in a trapping process. 
For simplicity, we assume that the radius of the trapped particle is much smaller than the wavelength of the beam ($R_0<20\lambda$) so that we can approximate 
the particle as an electric dipole, which is often referred to as the Rayleigh regime \cite{Harada1996529}. Here $\lambda$ and $R_0$ denote the wavelength of the 
optical beam and the particle radius, respectively. Using this approximation, the total electromagnetic force generated by the electric field ${\bf E}({\bf r})$
and the magnetic field  ${\bf B}({\bf r})$ in the presence of a dipole moment ${\bf p}$ can be expressed by  
\begin{equation}
\label{eq:refname1}
{\rm\bf F}({\rm\bf r})=({\rm\bf p}.{\nabla}){\rm\bf E}({\rm\bf r})+
\frac {\partial{{\rm\bf p}}} {\partial t}\times{\rm\bf B}({\rm\bf r})={\rm\bf F}_g({\rm\bf r})+{\rm\bf F}_s({\rm\bf r}),
\end{equation}
where ${\rm\bf F}_g({\rm\bf r})$ and ${\rm\bf F}_s({\rm\bf r})$ are the gradient and scattering forces, respectively.

The scattering force is mainly rooted in the reflections from the surface of the particle which result in a momentum recoil in the direction of 
the propagation of the beam, which is assumed to be in the $\hat{z}$ direction:
\begin{equation}
\label{eq:refname2}
{\bf F}_s(r)=\frac{n_2}{c} \sigma_s I({\bf r}) \hat{z},
\end{equation}
where $\sigma_s$ , $n_2$ and $c$ are the  scattering cross section, the refractive index of the surrounding medium and the speed of light, respectively,
and $I({\bf r})$ is the time-averaged intensity distribution at position ${\bf r}$. The scattering cross-section is given by
\begin{equation}
\label{eq:refname3}
\sigma_s=\frac{8}{3} \pi k_0^4 R^6_0  \Big(\frac{n_r^2-1}{n_r^2+2}\Big)^2,
\end{equation}
where $n_1$ is the refractive index of the particle, $n_r=n_1/n_2$ is the relative refractive index of the particle to that
of the surrounding medium, and $k_0=2\pi/\lambda$.

Nonuniformity in the intensity distribution of the optical field results in a non-vanishing gradient force (${\bf F}_g$) which always has the tendency to pull the 
particle to the high intensity region (if $n_r>1$). The gradient force can be expressed by
\begin{equation}
\label{eq:refname4}
{\bf F}_g(r)=\frac{2\pi n_2 R^3_0}{c} \Big(\frac{n_r^2-1}{n_r^2+2}\Big) \nabla I({\bf r}).
\end{equation}
Although the scattering force (${\bf F}_s$) always points in the direction of the beam propagation, 
the gradient force (${\bf F}_g$) can have any direction depending on the intensity profile of the optical beam. 
In particular, if the optical intensity at position ${\bf r}$ is a local maximum (minimum) for $n_r>1$ ($n_r<1$), 
then the position ${\bf r}$ is the minimum of the effective potential and is an optical trap~\cite{Ashkin:86}.
It must be noted that the existence of a local minimum of the optical potential is only a necessary condition for optical trapping and
other conditions need to be satisfied in order to obtain a stable optical trap as will be further discussed in section~\ref{sec:stability}.     
\section{Laguerre-Gaussian Beams}
In order to get a stable 3D optical trap, different optical beam profiles from zero-order Bessel to polarized beams have been 
employed~\cite{Meyrath:05,PhysRevLett.99.153603,:/content/aip/journal/apl/85/18/10.1063/1.1814820,PhysRevA.81.023831}. As it was 
mentioned earlier, we use a GIMF with the refractive index profile of the form 
\begin{equation}
\label{eq:refname5}
n^2(\rho)=n^2_{0}\Big[1-2\Delta(\frac{\rho}{R})^\alpha\Big],
\end{equation}
where $R$ is the core radius, $n_0$ is the refractive index at the center of the core, $\Delta$ is the index step, $\alpha\approx 2$ in 
the core ($\rho \le R$) and $\alpha=0$ in the cladding ($\rho\ge R$). The guided modes of the optical fiber are characterized by the
radial $p$ and angular $m$ integer numbers~\cite{Mafi:12}. The spatial profile of the nearly transverse electric fields corresponding to 
$p$ and $m$ are given in Ref.~\cite{Mafi:12}. The transverse electric field of each mode after propagation of a distance $z$ from the output 
tip of the fiber in the surrounding medium of refractive index $n_2$ is given by
\begin{align}
\label{eq:refname6}
&E_{p,m}(\rho,\phi,Z)=\sqrt{\frac{2 p!}{\pi (p+m)!}}\frac{1}{w(Z)} \Big(\frac{\sqrt{2}\rho}{w(Z)}\Big)^{|m|}\\
\nonumber
&\times\exp{\Big(\frac{-\rho^2}{w^2(Z)}\Big)}L_{p}^{|m|}\Big(\frac{2\rho^2}{w^2(Z)}\Big)
\exp{\Big(\frac{-ik{\rho}^2}{2R(Z)}\Big)}\\
\nonumber
&\times\exp{\big(im \phi\big)}\exp{\big(-ikz_0Z\big)} \exp{\big(ig\Psi(Z)\big)}.
\end{align}
$L_{p}^{|m|}$ are generalized Laguerre polynomials, $Z=z/z_0$, $z_0 =n_{2}\pi w^2_0/\lambda$ is the Rayleigh range, 
$k=n_2 k_0$, $\Psi(Z)=\tan^{-1}(Z)$ is the Gouy phase shift, and $g=2p+|m|+1$ is the group mode number. $R(Z)$ and $w(Z)$ 
are given by
\begin{align}
\nonumber
w^2(Z)&=w^2_{0} \Big(1+Z^2\Big),\\
R(Z)&=z_{0}Z\Big(1+Z^{-2}\Big).
\label{eq:refname7}
\end{align}

Throughout the manuscript, we assume that the electric filed is linearly polarized in the $\hat{x}$ direction; therefore, we
can use a scalar form for the amplitude of the electric field as presented in Eq.~\ref{eq:refname6}.
The modes in Eq.~\ref{eq:refname6} are normalized to unity according to 
\begin{align}
\int_0^{2\pi}d\phi\int^\infty_0d\rho~\rho~|E_{p,m}(\rho,\phi,Z)|^2=1,
\end{align}
and the fiber mode field profiles in Ref.~\cite{Mafi:12} are reproduced by setting $Z=0$ in Eq.~\ref{eq:refname6}
and using Snell refraction at the fiber tip~\cite{saleh2013fundamentals}.

In a GIMF, the propagation of the modes with different amplitudes, phases, and propagation constants results in
an interference pattern, which can propagate outside the fiber and ideally result in local intensity extrema
for trapping. The minimum number of modes that need to be excited in a GIMF in order to obtain trapping at a point 
away from the fiber tip is two. In the following, we will consider two scenarios: scenario $\mathcal{A}$ with only two lowest order 
modes, and scenario $\mathcal{B}$ with all the modes excited from an SMF similar to that of Ref.~\cite{Mafi:11}. In either scenario,
we only consider those modes with zero orbital angular momentum $m=0$. Those modes with $m\neq 0$ are interesting because
they can be used to apply torque and rotate particles, but our focus and interest are in manipulating the trapping position
in simplest experimental configurations, so $m=0$ consideration is justified. Due to the cylindrical symmetry, when SMF is center-spliced 
to the GIMF, only those modes with a zero orbital angular 
momentum number $m=0$ are excited. We note that both these scenarios are
well justified from a practical point of view as will be discussed in the appropriate subsections in the following.  
\section{Trapping with two spatial modes}
\label{Trapping with two spatial modes}
In this trapping scenario, we consider a GIMF where only the two lowest order modes with $m=0$ are excited. The two modes are 
considered here are $E_{0,0}$ and $E_{1,0}$. The intensity distribution outside the fiber can be generally expressed as
\begin{equation}
\label{eq:refname8}
I({\bf r})=\dfrac{1}{2}n_2 \epsilon_0 c~\Big({\dfrac{ \big|E_{0,0}+s \exp{\big(i\varphi\big)}E_{1,0}\big|^2}{1+s^2}}\Big).
\end{equation}
The parameter $s$ determines the relative amplitude of the $E_{1,0}$ to $E_{0,0}$ and $\varphi$ is the relative phase.
This two-mode scenario is justifiable in practical situations. In practice, an objective lens is used to couple the laser
light into the GIMF and in the absence of any major off-set or tilt, the $E_{0,0}$ mode is excited predominantly. In fact, 
the coupling laser light and the lens can be readily tuned to ensure that $E_{0,0}$ is nearly the only excited mode in the 
GIMF. Then, using a long period fiber grating (LPG)~\cite{Bhatia:96,Ramachandran:02}, it is possible to couple a portion of
$E_{0,0}$ to $E_{1,0}$. Therefore, the parameter $s$ is determined by the coupling strength through the amplitude and length
of the LPG, while $\varphi$ is determined by the difference between the propagation constants of the two modes and the
length of the GIMF. As such, this scenario is experimentally motivated.
  
The total intensity profile outside the fiber at the normalized distance $Z$ from the fiber tip is given by 
\begin{align}
\label{eq:refname9}
I({\bf r})=&\Big(\frac{n_2 \epsilon_0 c}{\pi}\Big) \Big(\frac{(1+s^2)^{-1}}{w^2(Z)}\Big)\exp{\big(\frac{-2\rho^2}{w^2(Z)}\big)}\\
\nonumber
&\times\Big|1+s \exp{\big(i\varphi\big)}\exp{\big(2i\Psi(Z)\big)}\Big(1-\frac{2\rho^2}{w^2(Z)}\Big)\Big|^2.
\end{align}  
Because of the cylindrical symmetry of the configuration, the trapping point for $m=0$ modes must lie on the z-axis. 
Therefore, we can set $\rho=0$ in Eq.~\ref{eq:refname9} in order to find the trapping location $Z$. 
After several algebraic steps, the intensity profile takes a simpler form
\begin{equation}
\label{eq:refname10}
I(0,\phi,Z)\propto{\Big(\frac{(1-a)Z^2-2bZ+(1+a)}{(1+Z^2)^2}\Big)},
\end{equation}
where 
$a=\sin\theta \cos\varphi$, $b=\sin\theta \sin\varphi$ and 
\begin{equation}
\theta:=2\tan^{-1}(s).
\label{eq:deftheta}
\end{equation}
Here, $\theta \in [-\pi,\pi]$ and $\varphi \in [-\pi,\pi]$.

The z-component of the gradient trapping force $F_g$ can be calculated using Eq.~\ref{eq:refname4}. The z-component of the gradiant of the intensity is given by 
\begin{equation}
\label{eq:refname11}
\frac{\partial I(0,\phi,Z)}{\partial Z}\propto -\dfrac{(1-a)Z^3-3bZ^2+(1+3a)Z+b}{\big(1+Z^2\big)^3}. 
\end{equation}
For later reference, we also calculate the normalized spring constant $k_{s,z}$ using the second derivative of the intensity profile,
which determines the stability of the trapping as will be discussed later 
\begin{align}
\label{eq:kz}
&k_{s,z}=\frac{\partial^{2} I(0,\phi,Z)}{\partial Z^{2}}\propto\\
\nonumber
&{\dfrac{3(1-a)Z^4-12 b Z^3+2(1+9a)Z^2+12 b Z-(1+3a)}{\big(1+Z^2\big)^4}}. 
\end{align}
\subsection{Trapping positions}
\label{Trapping positions}
In order to achieve trapping, we need to ensure that the gradient force is larger than the scattering force. Therefore,
calculating the trapping force using Eq.~\ref{eq:refname11} is merely a necessary condition for trapping and 
$F_g > F_s$ must also be verified. In a given fiber configuration and a laser wavelength, $F_g > F_s$ sets a maximum 
acceptable radius $R_0$ for the trapped particle as can be clearly seen in Eqs.~\ref{eq:refname2}, \ref{eq:refname3}, 
and~\ref{eq:refname4}, where $F_s\propto R_0^6$ and $F_g\propto R_0^3$.

In order to find the trapping position when $F_{s}$ is negligible compared to $F_{g}$, we only need to find
 the maximum of the intensity function by setting its derivative in Eq. \ref{eq:refname11} equal to zero. 
This results in a cubic equation of the form
\begin{equation}
\label{eq:refname13}
z^3+c_1 z^2+c_2 z+c_3=0
\end{equation}
where the coefficients are given by
\begin{subequations}
\begin{align}
c_1&=\frac{-3b}{1-a}z_0 = -(z_1+z_2+z_3)\\
c_2&=\frac{3a+1}{1-a}z^2_0=z_1z_2+z_1z_3+z_2z_3\\
c_3&=\frac{b}{1-a}z^3_0 =-(z_1 z_2 z_3).
\end{align}
\label{eq:Ccoefficeints}
\end{subequations}
$z_0$ is the Rayleigh range defined earlier. $z_1$, $z_2$, and $z_3$ are the roots of the cubic equation and are
related to $c_1$, $c_2$, and $c_2$ coefficients via Vieta's formulas~\cite{citeulike:9136246}.
A cubic equation with real coefficients can have three real roots or one real root and a complex conjugate pair. Here, we are only interested 
in real positive roots, which result in a trapping position outside and away from the fiber tip. In practice, it is preferrable to 
have a trapping position far enough from the fiber tip to reduce the possibility of the fiber tip contact with the 
particle (e.g. cells) and the potential damaging effects.

Vieta's formulas show that whenever a cubic equation has three positive roots, its coefficients should satisfy $c_1<0$ , $c_2>0$, and $c_3<0$. 
It is clear from Eq.~\ref{eq:Ccoefficeints} 
that these three conditions cannot be met simultaneously. Therefore, the maximum number of positive roots in this case is two. 
When $c_3<0$, there is only a single positive root. We also note that
this single positive root is always a stable point because $\partial^{2} I(0,\phi,Z)/\partial Z^{2}<0$ at this position. 
When $c_3>0$, there can either be two positive and one negative roots or 
two complex and one negative roots. We note that using Eq.~\ref{eq:Ccoefficeints} it can be shown that it is impossible to have 
three negative roots in this case. In order to constrain the problem to the case in which there are two positive and one negative roots, 
the following condition must be satisfied~\cite{irving2004integers}
\begin{equation}
\label{eq:refname17}
D=18c_1 c_2 c_3 -4 c_1^3c_3+c_1^2c_2^2-4c_2^3-27c_3^2\ge 0.
\end{equation}  
In general, if the discriminant ($D$) of the cubic equation becomes greater than and equal to zero, $D\ge0$, the cubic equation is bound to have 3 real roots. 
It is interesting to point out here that the bigger positive root is always stable and the smaller root is always unstable.

In the following, we will analyze the existence and positions of the roots using the two angular variables  $\varphi$ and $\theta$, which were defined 
in Eqs.~\ref{eq:refname8} and \ref{eq:deftheta}, respectively. In order to have $c_3>0$, these angles need to be in the ranges of 
$\varphi \in[0,\pi]$ and $\theta \in[0,\pi]$, or $\varphi \in[-\pi,0]$ and $\theta \in[-\pi,0]$. In order to have $c_3<0$, these angles need to 
be in the ranges of $\varphi \in[0,\pi]$ and $\theta \in[-\pi,0]$, or $\varphi \in[-\pi,0]$ and $\theta \in[0,\pi]$. However, a symmetry in
Eq.~\ref{eq:refname8} can simplify our analysis: this equation is invariant under the simultaneous transformations of the form $\varphi\to\varphi\pm\pi$
and $\theta\to -\theta$. Therefore, we only need to explore $\varphi \in[-\pi,\pi]$ and $\theta \in[-\pi,0]$, where $\varphi \in[-\pi,0]$ relates to
$c_3>0$ and $\varphi \in[0,\pi]$ relates to $c_3<0$.

\begin{figure}[htp]
  \centering
  \includegraphics[width=3.2in]{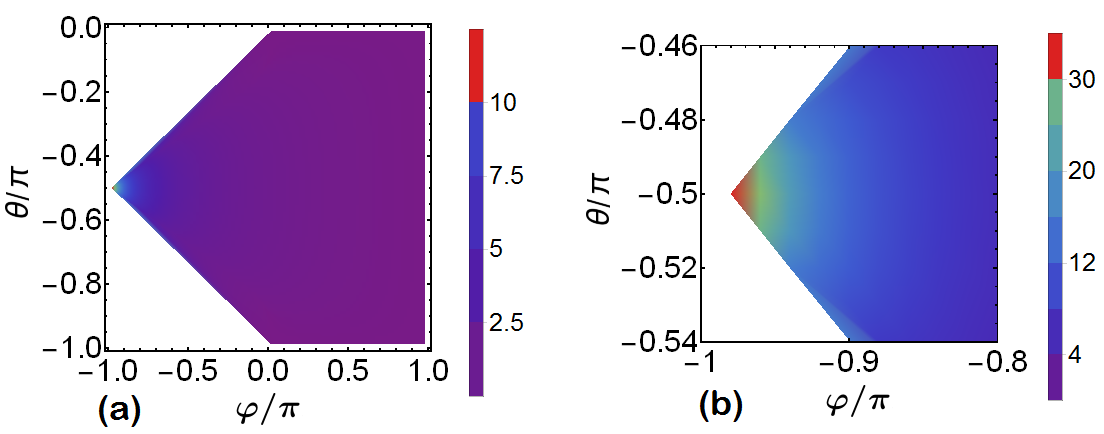}
\caption {(a) Normalized positions of the stable roots in the ranges of $\varphi \in[-\pi,\pi]$ and $\theta \in[-\pi,0]$; 
(b) zoomed-in image of normalized positions of the stable roots near the sharp region in Fig.~\ref{fig:roots}(a).}
\label{fig:roots}
\end{figure}
These observations are also supported by direct numerical simulations. In Fig.~\ref{fig:roots}(a), the root plot is presented over the relevant ranges 
of $\varphi$ and $\theta$. The colored regions correspond to the stable roots, while the white color corresponds to regions where 
the roots are always negative (located inside the fiber). The value in the density plot in Fig.~\ref{fig:roots}(a) represents the distance of the positive root from the tip of the fiber in units of $z_0$, i.e. $Z$. For example, Fig.~\ref{fig:roots}(a) shows that for $\varphi \in[0,\pi]$ and all values of $\theta \in[\pi,0]$, there is always 
a stable root, which is somewhat close to the tip of the fiber. However, for $\varphi \in[-\pi,0]$, the existence of a stable positive root depends on the 
value of $\theta$; e.g. $\varphi=\theta=-\pi$ admits no stable positive root. Moreover, stable roots that are quite far from the tip of the fiber and are of
practical interest occur near $\theta\approx -\pi/2$ and $\varphi\to -\pi^+$. Here $\to -\pi^+$ means approaching $-\pi$ from slightly larger values. This point 
corresponds to $se^{i\varphi}\approx 1$ so $E_{0,0}$ and $E_{1,0}$ modes contribute with equal amplitude and phase. Fig.~\ref{fig:roots}(b) is the same as 
Fig.~\ref{fig:roots}(a) except it is zoomed in near the sharp edge of the plot, in order to emphasize that the transitions, which look somewhat abrupt in 
Fig.~\ref{fig:roots}(a) are actually smooth and physically meaningful.

\begin{figure}[htbp]
  \centering
  \includegraphics[width=3.2in]{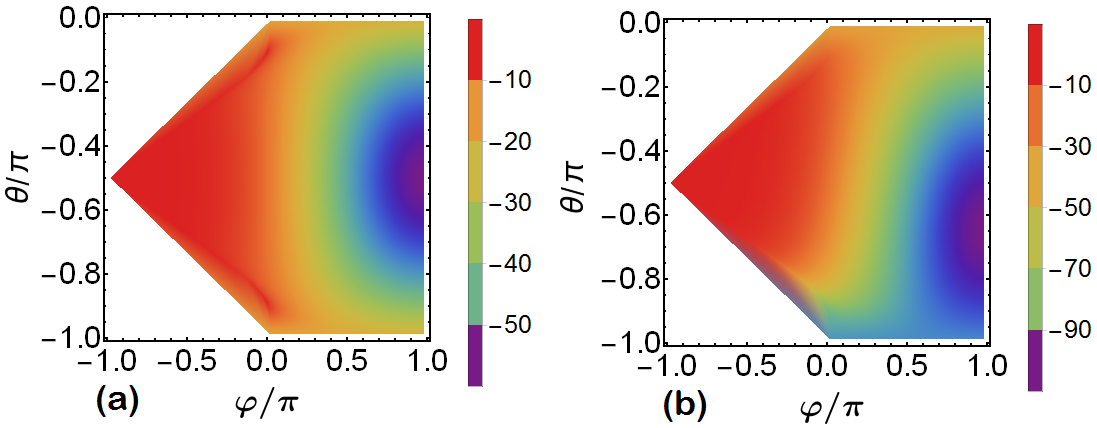}
\caption{(a) Relative longitudinal spring constant $k_{s,z}$ in the ranges of $\varphi \in[-\pi,\pi]$ and 
$\theta \in[-\pi,0]$ at the normalized positions of the stable roots; 
(b) relative transverse spring constant $k_{\rho,z}$ in the ranges of $\varphi \in[-\pi,\pi]$ and 
$\theta \in[-\pi,0]$ at the normalized positions of the stable roots.} 
\label{fig:spring-constants}
\end{figure}
The restoring force at the equilibrium point for a trapped particle determines the strength of the trapping and can be characterized by the spring constant, 
which is proportional to the second derivative of the optical intensity at the trapping location where its gradient vanishes. Of course, the spring constants
are different in the radial and longitudinal directions.
In Fig.~\ref{fig:spring-constants}(a) we plot the relative longitudinal spring constant $k_{s,z}=\partial^2 I/\partial Z^2$ at $\rho=0$ (see Eq.~\ref{eq:kz}) 
as a function of $\varphi$ and $\theta$ parameters in the arbitrary unit. It is clear that the point $\theta\approx -\pi/2$ and $\varphi\to -\pi^+$ corresponding 
to $se^{i\pi}\approx 1$ provides a long distance trapping position (Fig.~\ref{fig:roots}) and a weak longitudinal trapping (Fig.~\ref{fig:spring-constants}).
The maximum trapping longitudinal spring constant occurs at $\theta\approx -\pi/2$ and $\varphi\to \pi^-$. Here $\to \pi^-$ means approaching $\pi$ from slightly 
smaller values. This again corresponds   
to $se^{i\varphi}\approx 1$ so at this point $E_{0,0}$ and $E_{1,0}$ modes contribute with equal amplitude and phase as well. 

Fig.~\ref{fig:spring-constants}(b) shows the relative normalized transverse spring constant $k_{s,\rho}=(\partial^2 I(r)/\partial \rho^2+1/\rho~\partial I(r)/\partial \rho)$ 
at $\rho=0$ as a function of $\varphi$ and $\theta$ parameters. Fig.~\ref{fig:spring-constants}(b) shows that as the trapping position moves further away from the fiber tip, 
the transverse spring constant gets weaker like the longitudinal one.  
\section{Trapping with multiple excited modes from an SMF}
In practice, it may be difficult to only excite and work with two modes of a GIMF. A more practical approach was 
pursued in Refs.~\cite{Mafi:11,Hofmann:12}, where the laser was initially launched into an SMF, which was then
fusion spliced into a GIMF. Depending on the properties of the SMF and the GIMF, a few modes of the GIMF are excited
and propagate to the exit port of GIMF around which the trapping happens. In addition to the refractive index and mode-filed 
properties of the SMF and GIMF, the length of the GIMF also plays an important role in setting the trapping properties 
of the field as was also captured in parameter $\varphi$ of Eq.~\ref{eq:refname8} for the two-mode scenario in the previous section.   
Unlike other multimode optical fibers, GIMF has a very short self-imaging length typically less than 1~mm, which can be easily 
adjusted by simple polishing. Therefore, the entire fiber-length-related parameter space is accessible easily by merely polishing
the fiber tip, which is of immense practical importance.

\begin{figure}[htbp]
  \centering
  \includegraphics[width=3.2in]{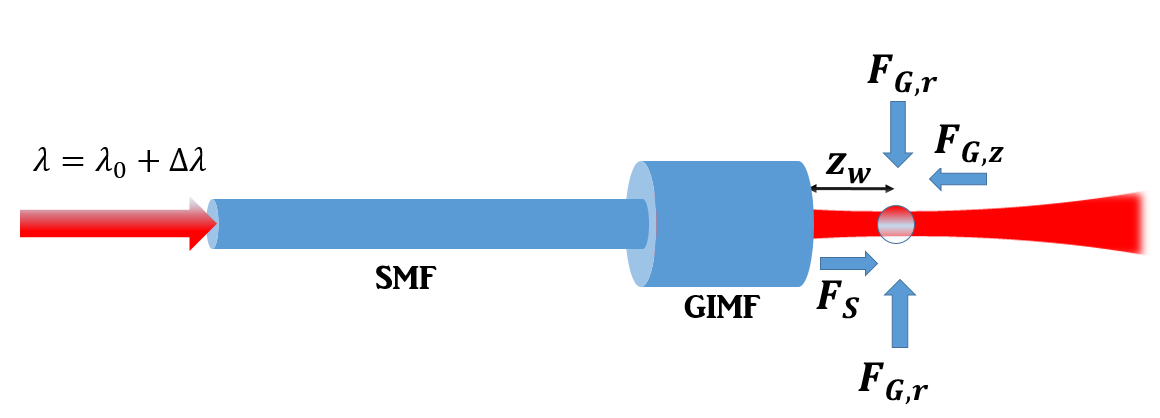}
\caption{Schematic of the proposed OFT.}
\label{fig:tweezerpic1}
\end{figure}
Consider the nearly Gaussian beam of an SMF injected into the GIMF as depicted in Fig.~\ref{fig:tweezerpic1} similar to Ref.~\cite{Mafi:11}.
The input electric beam profile can be expressed as
\begin{equation}
\label{eq:refname18}
E_{i}(\rho)=E(0,0)\exp{\big(-\frac{\rho^2}{\Omega^2}\big)}.
\end{equation} 
We have
\begin{equation}
\label{eq:refname19}
E(0,0)=2\sqrt{\frac{P_{0}\eta}{\pi {\Omega}^2}},\qquad \eta=\sqrt{\dfrac{\mu_0}{\epsilon_{\rm SMF}}},
\end{equation}
where $P_0$ is the input power, $\Omega$ is the input beam width, and $\eta$ and $\epsilon_{\rm SMF}$ are the average impedance and permitivity of the SMF, respectively. 

It can be shown analytically that the input electric beam profile can be expanded in terms of the guided modes of the GIMF as shown below~\cite{Mafi:11}
\begin{align}
\label{eq:refname20}
E_{i}(\rho)&=2\sqrt{\frac{\eta P_{0}}{\pi {\Omega}^2}}\exp{\big(-\frac{\rho^2}{\Omega^2}\big)}\\   \nonumber
&\approx 4\sqrt{\frac{P_{0}\eta}{\pi {\Omega}^2}} \dfrac{1}{\zeta+1} \exp{\big(-\frac{\rho^2}{w_{0}^2}\big)} \sum_{p=0}^{N-1} L_{p}^{0}\Big(\frac{2\rho^2}{w^2_{0}}\Big)\gamma^{p}.
\end{align}
We have defined
\begin{subequations}
\begin{align}
\label{eq:refname21}
\zeta&=\dfrac{w_{0}^2}{\Omega^2},\\
\label{eq:refname21b}
\gamma&=\frac{\zeta-1}{\zeta+1},\\
\label{eq:refname21c}
w_0&=\dfrac{(R \lambda)^{\frac{1}{2}}}{(\pi n_0)^{\frac{1}{2}}(2\Delta)^{\frac{1}{4}}},
\end{align}
\end{subequations}
where $w_0$ is the width parameter of Laguerre-Gaussian modes inside the fiber, $\gamma$ is the coupling coefficient of the input electric beam profile
to the guided modes inside the fiber, $n_0$ is the refractive index of the core of the GIMF, $R$ is the GIMF core radius, $\Delta$ is the index step, 
and $N$ is the total number of the guided modes within the fiber.

The output electric profile $E(\rho,z)$ at the tip of the fiber after the propagation of $L$ would be
\begin{align}
\label{eq:refname23}
E(\rho,z)\Big{|}_{z=L}&= 4\sqrt{\frac{P_{0}\eta}{\pi {\Omega}^2}} \dfrac{\exp{(i\beta_1 L)}}{\zeta+1}\\
\nonumber
 &\times\exp{\big(-\frac{\rho^2}{w_{0}^2}\big)} \sum_{p=0}^{N-1} L_{p}^{0}
\Big(\frac{2\rho^2}{w^2_{0}}\Big)\gamma^{p}\exp^{\big(2i\mathcal{Z}p \big)},
\end{align}
where we have
\begin{subequations}
\begin{align}
\label{eq:refname24}
\mathcal{Z}&=\frac{\sqrt{2\Delta}}{R} L,\\
\beta_{1}&= n_{0} k_{0}-\dfrac{\sqrt{2\Delta}}{R}.
\end{align}
\end{subequations}

It is important to mention that due to the cylindrical symmetry, when SMF is center-spliced to the GIMF, only those modes with a zero orbital angular 
momentum number $m=0$ are excited. These modes inevitably have an odd mode group number $g$. This means that the total number of propagating modes in a
GIMF in this configuration will be quite less than the total number of modes, which can be supported in a GIMF. Inserting the electric field profile at 
the fiber tip $E(\rho,L)$ into the Generalized Huygens integral, the propagating electric field in the medium surrounding the GIMF can be obtained. 
Due to the fact that Laguerre-Gaussian functions are self-transforming under the Hankel transformation, the electric field profile function preserves 
its own Laguerre-Gaussian form under propagation~\cite{Tache:87}. The propagating electric field profile in the surrounding medium can be expressed by
\begin{align}
\label{eq:refname25}
E(\rho,Z)&= 4\sqrt{\frac{P_{0}\eta}{\pi {\Omega}^2}}\dfrac{\exp{(i\beta_1 L)}}{\zeta+1} \frac{w_0}{w(Z)}\exp{\big(\frac{-\rho^2}{w^{2}(Z)}\big)}\\
\nonumber
&\times \sum_{p=0}^{N-1}  L_{p}^{0}\Big(\frac{2\rho^2}{w^2(Z)}\Big) \big(\gamma^p e^{2i\mathcal{Z}p}  e^{i (2p+1)\Psi(Z)}\big).
\end{align}
After some simplifications, the time-averaged intensity profile along the $z$ axis takes the form of 
\begin{align}
\label{eq:refname27}
I(0,Z)&=\dfrac{8P_0 }{\pi \Omega^2}\big(\dfrac{n_2}{n_{\rm SMF}}\big)\dfrac{1}{(1+\zeta)^2} \frac{1}{1+Z^2}\\
\nonumber
&\times\frac{1+\gamma^{2N} -2 \gamma^{N}\cos\big(2N(\mathcal{Z}+\Psi(Z))\big)}{1+\gamma^2-2\gamma\cos\big(2(\mathcal{Z}+\Psi(Z))\big)},
\end{align}
where $n_{\rm SMF}$ is the refractive index of the SMF. It is easy to verify that by setting $N=2$ in Eq.~\ref{eq:refname27}, we obtain Eq.~\ref{eq:refname10}
if the normalization coefficient $1+\gamma^2 $ is properly taken into account. It must be noted that $2\mathcal{Z}$ and $\gamma$ of Eq.~\ref{eq:refname27}
play the roles of $\varphi$ and $s$ in the Eq.~\ref{eq:refname10}, respectively.

Equation~\ref{eq:refname27} can be compared with the derivation for a system consisting of a Graded-index (GRIN) lens with an incident Gaussian beam injected through an SMF, which is similar to the
SMF launching beam profile considered here. The equation for GRIN lens can be obtained as the $N\to\infty$ limit of Eq.~\ref{eq:refname27}, which reduces to
\begin{align}
\label{eq:refname28}
I(0,Z)\Big|_{N\to\infty}\approx & \dfrac{8 P_0}{\pi \Omega^2}\big(\dfrac{n_2}{n_{\rm SMF}}\big)\dfrac{1}{(1+\zeta)^2} \frac{1}{1+Z^2}\\
\nonumber
&\times \frac{1}{1+\gamma^2-2\gamma\cos\big(2(\mathcal{Z}+\Psi(Z))\big)}.
\end{align}
The intensity distribution in Eq.~\ref{eq:refname28} peaks at a distance $z_w$ from the lens tip, which is effectively the working distance for the GRIN lens 
as also derived previously by Jung et. al.~\cite{jung2010numerical} using the matrix optics technique: 
\begin{equation}
\label{eq:refname29}
z_w=-z_0 \Big[\dfrac{2\gamma \sin(2\mathcal{Z})}{1+\gamma^2+2\gamma\cos(2\mathcal{Z})}\Big].
\end{equation} 

Using Eq.~\ref{eq:refname27}, we can predict the trapping location given the geometrical and optical properties of the SMF and GIMF fibers.
Here, we consider two different setups using a set of GIMFs in Table~\ref{tab:table1} and SMFs in Table~\ref{tab:table2}.
GIF625 is by Thorlabs Inc., GC.400/500 is by Fujikura Ltd., and UHNA3 is by Nufern Inc. The relevant parameters are all given in
Tables~\ref{tab:table1} and \ref{tab:table2}.

\begin{table}
\caption{\label{tab:table1} Characterizations of the different used GIMFs at the wavelength of $\lambda=1.55~\mu m$.}
\renewcommand{\arraystretch}{1.2}
\begin{ruledtabular}
\begin{tabular}{ |p{2cm}|p{1.3cm}|p{1.3cm}|p{1.5cm}|p{1.1cm}| } 
fiber type & core diameter ($\mu m$) & cladding diameter ($\mu m$) & Numerical aperture (NA) & MFD ($\mu m$)\\
\hline
GIF625 & 62.5 & 125 & 0.275 & 15.2\\
GC.400/500 & 400 & 500 & 0.2 & 42\\
\end{tabular}
\end{ruledtabular}
\end{table}  
\begin{table}
\caption{\label{tab:table2} Characterizations of the used SMF at the wavelength of $\lambda=1.55~\mu m$.}
\begin{ruledtabular}
\renewcommand{\arraystretch}{1.2}
\begin{tabular}{ |p{2cm}|p{1.3cm}|p{1.3cm}|p{1.5cm}|p{1.1cm}| } 
fiber type & core diameter ($\mu m$) & cladding diameter ($\mu m$) & Numerical aperture (NA) & MFD ($\mu m$)\\
\hline
UHNA3 & 1.8 & 125 & 0.35 & 4.1\\
\end{tabular}
\end{ruledtabular}
\end{table}
We consider two scenarios: {\em Case 1} where the SMF ``UHNA3'' is connected to the GIMF ``GIF625'', 
and {\em Case 2} where the SMF ``UHNA3'' is connected to the GIMF ``GC.400/500''. In order to use
Eq.~\ref{eq:refname27} to calculate the intensity distribution for the fiber tweezer, we also need to 
know the total number of propagating modes  $N$ in the GIMF in each case. Equation~\ref{eq:refname30a}
gives the maximum mode group number $g_N$ in a GIMF. 
\begin{align}
\label{eq:refname30a}
g_N\approx {\rm NA}\times(\frac{\pi R}{\lambda}).
\end{align}
Because $g_N=2p+|m|+1$ for values of $p$ and $m$ belonging to the highest order mode group, and because we 
only consider $m=0$ modes, we can use $g_N=2N+1$. Therefore, we have
\begin{align}
\label{eq:refname30b}
2N+1 \approx {\rm NA}\times(\frac{\pi R}{\lambda}),
\end{align}
which can be solved for $N$ to obtain the total number of $m=0$ propagating modes here.

Using the necessary information in the Tables~\ref{tab:table1} and \ref{tab:table2} and Eqs.~\ref{eq:refname21b}
and~\ref{eq:refname30b}, we obtain ($N=8$, $\gamma=0.86$) and ($N=42$, $\gamma=0.98$) for {\em Case 1} and {\em Case 2} scenarios, respectively. 
We note that GC.400/500 can be considered as a GRIN lens because it supports a very large number of propagating modes. Although GC.400/500 
can be treated like a GRIN lens, it is not an ideal choice as a tweezer because of its low numerical aperture~\cite{K.Tagucht}.

\begin{figure}[t]
  \centering
  \includegraphics[width=3.4in]{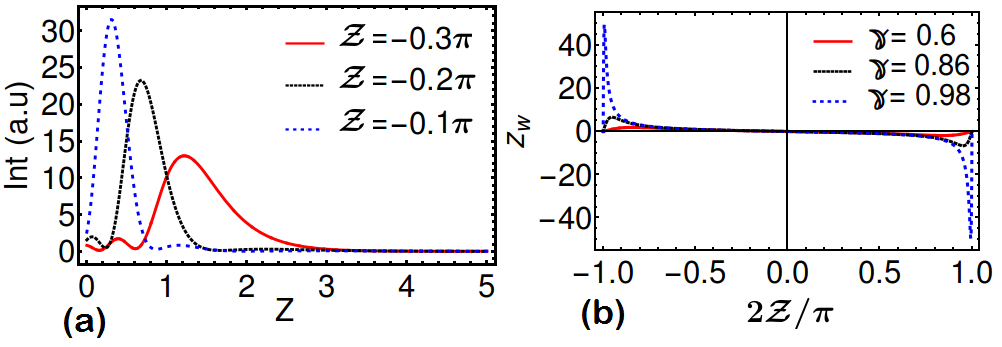}
\caption{(a) Intensity distribution in arbitrary unit (a.u) for {\em Case 1} at the center line ($\rho=0$) as a function of the normalized 
distance $Z$ for different values of $\mathcal{Z}$; 
(b) normalized working distance $Z_w$ plotted as function of $\mathcal{Z}$ for three different coupling coefficients 
$\gamma$=$0.6$, $0.86$, $0.98$ at $\lambda$=$1.55~\mu m$.}
\label{fig:refname5}
\end{figure}
Fig.~\ref{fig:refname5}(a) shows the intensity distribution at the center line ($\rho=0$) as a function of the normalized distance $Z$ from the tip of
the GIMF for {\em Case 1}. Different curves correspond to different values of the normalized length $\mathcal{Z}$ of the GIMF. It is observed 
that the value and location of the maximum intensity changes with the value of $\mathcal{Z}$, which is in agreement with the
observations of Zhang et. al.~\cite{Zhang} who showed that the tweezing location changes in the presence of the fluidic forces , when the GIMF is stretched under
tension. 

In Fig.~\ref{fig:refname5}(b) the normalized working distance $Z_w=z_w/z_0$ of the GIMF tweezer is plotted as a function of 
the normalized GIMF length $\mathcal{Z}$ for different values of $\gamma$. It is shown that the working distance of the tweezer 
system is strongly dependent on the value of $\gamma$: the main peak shifts to larger distances from the fiber tip when $\gamma$ increases.
We remind that {\em Case 1} corresponds to $\gamma=0.86$, while {\em Case 2} corresponds to $\gamma=0.98$. Therefore, the trapping location
is quite far from the tip of the fiber in {\em Case 2}. 
\section{Wavelength tunability}
The Ge-doped silica core and the undoped silica cladding of the GIMF have different refractive indices and dispersion properties as discussed
in Refs.~\cite{Mafi:11,Hofmann:12}. The dispersive behavior of the parameter $\Delta$ used in  Eq.~\ref{eq:refname24} is shown in Ref.~\cite{Mafi:11}
to follow the following form:
\begin{equation}
\label{eq:refname31}
\Delta=\Delta_0+\delta (\lambda-\lambda_0), 
\end{equation} 
where using the data from Hofmann et al.~\cite{Hofmann:12}, we obtain 
$\delta\approx 0.35/mm$ and $\Delta_{0}=0.02$ for the GIF625 at $\lambda_0=1.55~\mu m$. 
Due to the wavelength dependence of $\Delta$ and based on Eq. \ref{eq:refname29}, the 
confocal parameter of the GIMF  $2z_0$  and the normalized length $\mathcal{Z}$ which establish the base phase difference 
between the propagating modes in Eq.~\ref{eq:refname25} become dependent on the wavelength. 
Therefore, a change in the wavelength results in a change in the intensity distribution (Eq.~\ref{eq:refname27}). 
The dependence of $\mathcal{Z}$ on the wavelength $\lambda$ can be expressed as 
\begin{subequations}
\begin{align}
\label{eq:refname32}
\mathcal{Z}&\approx \mathcal{Z}\big|_{\lambda_0}+\mathcal{Z}(\lambda)\\
\label{eq:refname321}
\mathcal{Z}(\lambda)&= \frac{\delta}{\sqrt{2\Delta_0}} \big(\frac{L}{R}\big) (\lambda-\lambda_0).    
\end{align}
\end{subequations}

\begin{figure}[htbp]
  \centering
  \includegraphics[width=3.2in]{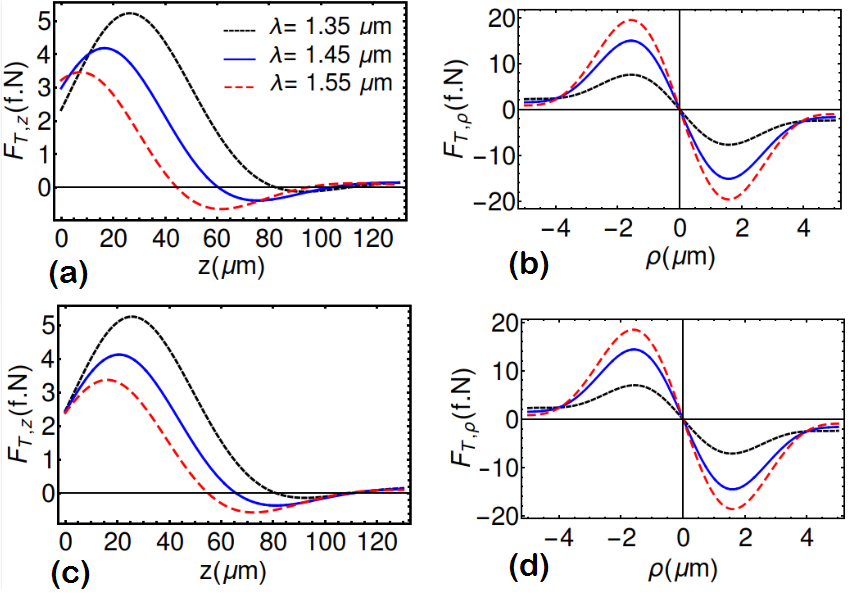}
\caption{Total radiation forces in femto-Newtons (f.N) for two different lengths of GIMF $L=9.3~mm,3.4~mm$ and at three different wavelengths
$\lambda=1.35\mu m$,$1.45~\mu m$, $1.55~\mu m$: (a) total axial radiation force ($F_{T,z}$) with the GIMF length of $L=9.3~mm$; (b) total radial 
radiation force ($F_{\rho,z}$) with the GIMF length of $L=9.3mm$; (c) total axial radiation force with the GIMF length of $L=3.4~mm$; 
(d) total radial radiation force with the GIMF length of $L=3.4~mm$.}
\label{fig:refname6}
\end{figure}

 Besides the dispersive effects that lead to the wavelength dependence of the NA of the GIMFs as it was explained above, 
the beam width $w_0$ exiting the GIMF is also wavelength dependent, resulting in the wavelength dependence of the  coupling coefficient $\gamma(\lambda)$. 
For example for {\em Case 1} we have $\gamma(1.35~\mu m)=0.84$, $\gamma(1.45~\mu m)=0.85$, and $\gamma(1.55~\mu m)=0.86$. 
Because the intensity distribution is a function of the coupling coefficient (Eq.~\ref{eq:refname27}), changing the coupling coefficient results in a change in the trapping position.   
The impact of the wavelength from both the normalized length $\mathcal{Z}$ and the coupling coefficient $ \gamma(\lambda)$ can be seen in Fig.~\ref{fig:refname6}
where the total force distribution $F_T$ exerted on a particle with a radius of $R_0=100~nm$ at the center line ($\rho=0$) is shown 
for $\lambda=1.35$~$\mu m$, $\lambda=1.45$~$\mu m$, and $\lambda=1.55$~$\mu m$, and GIMF lengths of 9.3~mm and 3.4~mm. All scenarios are related to {\em Case 1},
where the UNHA$3$ is pumped with an input power of $P_0=400~mW$ and immersion liquid is also water ($n_2=1.33$). Fig.~\ref{fig:refname6} (a) corresponds to a 9.3~mm GIMF and shows that as the wavelength changes, the trapping position changes. Figure~\ref{fig:refname6}(c) corresponds to a 3.4~mm GIMF and it is clear that the wavelength dependence of the trapping position
strongly depends on the length of the GIMF. We remind that because of the self-imaging behavior, all our observations would be the same for a GIMF length of
$L$ and $L+\pi R/\sqrt{2\Delta}$. Figs.~\ref{fig:refname6}(b) and (d) show that by changing the wavelength, the radial stability condition for the 3D optical 
trap remains the same on the axis.    
\section{Stability Condition}
\label{sec:stability}
In order to have a stable 3D optical trap, at least two conditions have to be met~\cite{Harada1996529,Liu:12}. The first one which is a necessary condition can be expressed 
as $F_G>F_S$. This condition is commonly expressed using the ratio of the forces~\cite{Harada1996529,Liu:12} 
\begin{equation}
\label{eq:refname33}
R_F=\Big|\dfrac{F_g}{F_s}\Big|\ge1  .  
\end{equation}
We already mentioned in section~\ref{Trapping positions} that this condition sets a maximum acceptable radius $R_0$ for the trapped particle.
The second necessary condition is that the trapping potential $U_{\rm trap}$ must be larger than $kT$, where $k$ is the Boltzmann constant and $T$ is the temperature
of the system~\cite{Liu:12}. This condition ensures that the thermal fluctuations do not drive the trapped particle out of equilibrium in the optical trap. This condition can be expressed as
\begin{subequations}
\begin{align}
\label{eq:refname34}
R_{th}&=\exp\Big(-\frac{U_{\rm trap}}{kT}\Big)\ll1,\\
\label{eq:refname35}
U_{\rm trap}&=\pi \epsilon_0 n_2^2 R^3_0 \Big|\dfrac{n^2_r-1}{n^2_r+2}\Big||E_{max}|^2,
\end{align}
\end{subequations}
where $|E_{max}|$ is the maximum value of the electric field profile. This condition can be met by increasing the total optical power
if the medium does not heat up significantly. In the presence of a fixed optical power, this condition translates into a minimum value for the radius of the trapped particle.

\begin{figure}[htbp]
  \centering
  \includegraphics[width=3.2in]{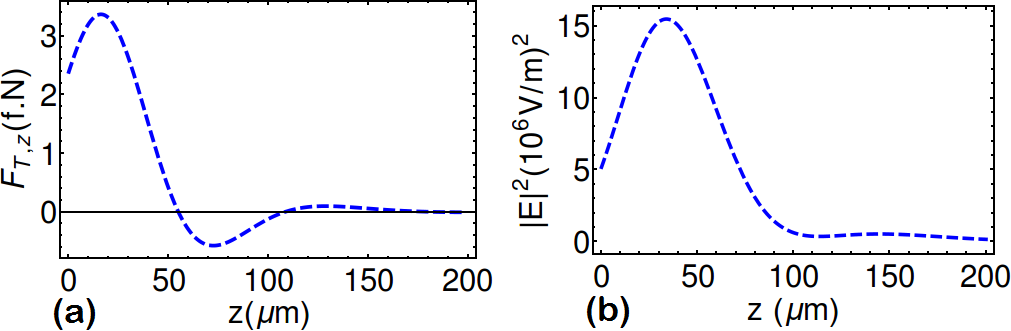}
\caption{(a) Total axial radiation force ($F_{T,z}$) over the $z$ axis for $\lambda=1.55~\mu m$ and $L=3.4~mm$; (b) the square absolute value of the 
electric field distribution ($|E(z)|^2$) over the $z$ axis for $\lambda=1.55~\mu m$ and $L=3.4~mm$.}
\label{fig:refname7}
\end{figure}

As we said above, in order to satisfy Eq.~\ref{eq:refname34}, the electric potential energy $U$ should be high enough ($U_{trap}\ge10 kT$) to dominate the thermal energy. 
To investigate the stability condition, we consider again {\em Case 1} with an input power of $P_0=400~mW$. We also consider polystyrene beads with the radius of $R_0=100$~nm and refractive index of $n_1=1.59$ which are surrounded by water. It is important to note that so far we have used the dipole approximation or Rayleigh condition that is allowed only when the wavelength is much larger than the particle radius $(\lambda/R_0)\ge 20 $, otherwise we should resort to the other methods such as Generalized Lorenz-Mie theory to calculate the scattering force~\cite{Harada1996529}. Here $\lambda/R_0\approx 16$ which is not far from the ideal ratio of 20 and still gives us a reliable stability condition~\cite{Harada1996529}. 
Fig.~\ref{fig:refname7}(a) shows the total trapping force distribution on the central line as a functions of the separation from the fiber tip in this scenario when the 
length of the GIMF and the wavelength are $L=3.4~mm$ and $\lambda=1.55~\mu m$ respectively. The figure shows that there is a trapping position where the total force changes sign. The presence of the trapping position tells us that the gradient force is larger than scattering force hence satisfying the first stability condition. Fig.~\ref{fig:refname7}(b) expresses the square absolute value of the electric field on the central line as calculated from Eq.~\ref{eq:refname25} for the input power of $P_0=400~mW$ as a function of the separation from the fiber tip. Using the stability condition of $U_{\rm trap}>10 kT$, we obtain a minimum required radius of $R_0=78~nm$, which is smaller than the particle radius selected in this example. Therefore, beads with the radius of $R_0=100~nm$ should have no problem satisfying the thermal stability condition.
\section{Conclusion}
We have proposed a wavelength-tunable OFT based on GIMF with a flat endface and without the use of any fluidic forces. 
We consider two scenarios: one in which only two propagating modes in the fiber are excited; and another 
where multiple modes are excited by butt-coupling a single-mode fiber to the GIMF. Our analytical calculations have 
shown that the trapping position of the proposed 
tweezer can be manipulated over a long distance from the fiber tip by tuning the wavelength. Moreover,
it is shown that the changes in the structural geometries of the GIMF such as the length of the GIMF can alter the trapping position.
Our analyses for the stability conditions of the proposed OFT implies that a 3D optical trap for a polystyrene bead with the radius 
of $R=100~nm$ is achievable using a 400~mW laser.
\bibliography{blist}
\end{document}